\documentclass[hyperref,amsmath,amssymb,showpacs,floatfix,journal=ancac3,manuscript=article]{achemso}
\usepackage{graphicx}	
\usepackage{amsmath}	
\usepackage{amssymb}	
\usepackage{extarrows}
\usepackage{morefloats}
\mciteErrorOnUnknownfalse 
\usepackage[version=4]{mhchem} 
\setkeys{acs}{etalmode=truncate, maxauthors=200}
\setkeys{acs}{keywords = true}
\setkeys{acs}{usetitle=true}
\usepackage{epstopdf} 
\usepackage[perpage]{footmisc}

\usepackage[final]{changes}
\DeclareGraphicsExtensions{.eps, .ps, .pdf}
\graphicspath{{./}{/home/ioan/floppy/astro/mnras/}{/home/ioan/floppy/astro/mnras/data/}{/home/ioan/floppy/astro/mnras/data_c4n-/}}

\newcommand{\gl}{equation} 
\newcommand{\gls}{equations} 
\newcommand{\Gl}{Equation}
 
\newcommand{\figname}{Figure~} 
\newcommand{\figsname}{Figures~}
\newcommand{\secname}{Section~}

\usepackage{mfirstuc}
\MFUnocap{of}
\MFUnocap{that}
\MFUnocap{by}
\MFUnocap{to}
\MFUnocap{on}
\MFUnocap{and}
\MFUnocap{in}
\MFUnocap{the}
\MFUnocap{for}
\MFUnocap{with}
\SectionNumbersOn 
\author{Ioan B\^aldea}
\email{ioan.baldea@pci.uni-heidelberg.de}
\affiliation{Theoretische Chemie, Universit\"at Heidelberg, Im Neuenheimer Feld 229, D-69120 Heidelberg, Germany}
\title[Suppression of Groups Intermingling as Appealing Option For Flattening and Delaying the Epidemiological Curve While Allowing Economic and Social Life at Bearable Level During COVID-19 Pandemic]  
{Suppression of Groups Intermingling as Appealing Option For Flattening and Delaying the Epidemiological Curve While Allowing Economic and Social Life at Bearable Level During COVID-19 Pandemic}

\begin{document}

\begin{abstract}
  In this work, we simulate the COVID-19 pandemic dynamics in a population modeled as a
  network of groups wherein infection can propagate both via intra-group and via inter-group
  interactions.
  Our results emphasize the importance of diminishing
  the inter-group infections in  
  the effort of substantial flattening and delaying of the epi(demiologic) curve with concomitant
  mitigation of disastrous economy and social consequences.
  To exemplify with a limiting case, splitting a population into m (say, 5 or 10)
  noninteracting groups while keeping intra-group interaction unchanged yields 
  a stretched epidemiological curve having the maximum number of daily infections
  reduced and postponed in time by the same factor $m$ (5 or 10).
  More generally, our study suggests a practical approach to fight against
  SARS-CoV-2 virus spread
  based on population splitting into groups and minimizing intermingling between them.
  This strategy can be pursued by large-scale infrastructure reorganization of activity 
  at different levels in big logistic units (e.g., large productive networks,
  factories, enterprises, warehouses, schools, (seasonal) harvest work). 
  Importantly, unlike total lockdwon strategy,
  the proposed approach prevents economic ruin and keeps social life at a more bearable level
  than distancing everyone from anyone.
  Last but not least, the declaration, for the first time in Europe, that COVID-19 epidemic ended in
the two million people Slovenia may be taken as confirming the advantage of the strategy
proposed in this paper.
 \end{abstract}
\section{Introduction}
\label{sec:intro}
Neither natural immunization nor vaccination or pharmacologic intervention can currently help
in fighting the COVID-19 global pandemic \cite{WHO_COVID-19_Pandemic}.
Most frequently, to prevent disastrous sanitary consequences, governments across the world responded
by imposing, e.g., social (=physical) distancing,
wearing face masks, lockdown regulations, and rigid sanitary moves
aiming at reducing the infection rate (``flattening the epidemiological curve'').
Still, politics cannot push strict restrictions indefinitely,
and ``how much is too much?'' is a question of time which unavoidably arises sooner or later.
Fighting COVID-19 should not ruin economy \cite{Baldwin:20}. This is certainly what a
draconian lockdown across the world over months would do. Parenthetically,
catastrophic economic consequences inherently make healthcare system itself also collapsing.

In this vein, mathematical
modeling may make a notable contribution in providing politics with reasonable suggestions 
to slowing down epidemic propagation and reducing medical burden while mitigating economy
and social crisis. 
A series of mathematical COVID-19 simulations have recently appeared
\cite{Chen2020.01.19.911669,Su2020.03.06.20032177,Li:2020,Roosa:2020,Shen2020.01.23.916726,8b30d75ac97f46a0bafcad5166faa39f,Yang2020.04.01.20043794,aggarwal2020.04.17.20069245}.
Most of them are based on deterministic continuous-time
epidemiological models, which consider age-independent
epidemiological classes of, e.g., susceptible (S), exposed (E), infected (I), and recovered (R) individuals
\cite{Kermack:27,Kermack:32,Kermack:33,Bailey:75,Hethcote:94,Hethcote:00}, whose numbers $S(t)$, $E(t)$,
$I(t)$, $R(t)$ evolve in time ($t$) according to a system of (deterministic) ordinary differential equations.

Open access sources already available \cite{SEIR_simulator,MathematicaSEIR,EpidemicCalculator} enable one to easily
perform various numerical simulations by means of such models.
Unfortunately the various SIR-inspired flavors need (too many) input parameters difficult
to validate \cite{IJIMAI-3841},
and this would rather mask than enlighten the main idea which the present work aims at conveying.
Therefore, to better emphasize this idea, instead of a SIR-based approach (which would pose no special problem),
in this paper we prefer to adopt the simpler logistic growth framework.
The logistic model is particularly appealing in view of its simplicity and versatility demonstrated in approaching 
a broad variety of real systems with very different nature 
\cite{Quetelet:1848,Ostwald:1883,McKendrick:1912,Lloyd:67,Cramer:02,Vandermeer:10,Baldea:2017m,Baldea:2018e},
including population dynamics of epidemic states \cite{Mansfield:60,Waggoner:00,Koopman:04,Bangert:17}.
Prior to this study, results based on the logistic model were presented  
for COVID-19 time evolution in China and USA \cite{Hermanowicz2020.03.31.20049486}. 

Fighting against the spread of SARS-CoV-2 virus while allowing economic and social
activity to continue to a reasonable extent represents a major challenge for the present era.
Extended lockdown does not represent an acceptable response to this challenge.
From this perspective, we believe that the results reported below obtained by extending the conventional logistic model
may provide useful suggestions on how 
to sidestep the ongoing difficulty of living under pandemic conditions.

While the implementation of the presently proposed strategy 
via population splitting into smaller groups and reducing intermingling
certainly requires considerable effort and fantasy in infrastructure reorganization,
it offers the perspective of flattening and delaying the epidemiological curve
by obviating wrecking of economy and maintaining social life
to a level more bearable than total lockdown.
%
\section{Methods}
\label{sec:method}
The results reported below were obtained by means of the logistic model
\cite{Quetelet:1848,Ostwald:1883,McKendrick:1912,Lloyd:67,Cramer:02,Vandermeer:10,Baldea:2017m,Baldea:2018e}
extended (see \gl~(\ref{eq-general-coupling}))
to allow treatment of infection propagation in a network consisting of
groups in interaction.
To fix the ideas and to make the paper self-contained, in \secname\ref{sec:model} 
we will first review the main aspects related to the logistic model applied to an isolated group
using a terminology adapted to the specific subject under consideration. The extension of the
logistic model to groups in interaction will be presented in \secname\ref{sec:groups}.
\section{Results and Discussion}
\label{sec:results}
\subsection{Logistic Growth in an Isolated Group}
\label{sec:model}
\begin{figure}
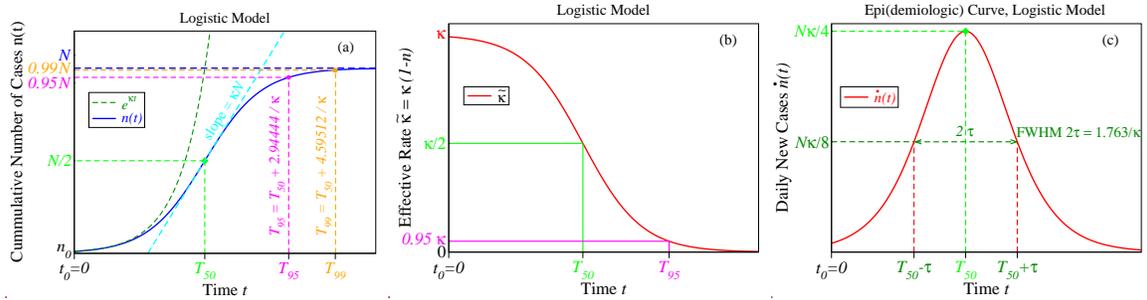

  \centerline{\includegraphics[width=0.3\textwidth]{fig_f_k_2}
    \includegraphics[width=0.3\textwidth]{fig_effective_rate_k_2}
    \includegraphics[width=0.3\textwidth]{fig_df_k_2}
  }
  \caption{
    Logistic growth is characterized by a cumulative number of infected cases $n(t$) following an S-shaped (sigmoid) curve
    (panel a) exhibiting an acceleration stage (close to a J-shaped exponential growth), which switches to a deceleration stage
    beyond the half-time $T_{50}$ and attains $p$($=95, 99$, see \gl~(\ref{eq-Tp}))
    percent of the plateau value $N$.
    Saturation occurs because the effective infection rate $\tilde{\kappa}$ is time dependent and
    gradually decreases to zero (panel b).
    The epi(demiologic) curve $\dot{n}(t) \equiv d n/d t$
    (daily number of new cases) exhibits a peak located at $t=T_{50}$
    whose shape is controlled by the infection rate $\kappa$ (panel c).
    Time on the $x$-axis is expressed in units of the characteristic time
    $T_{c} =\kappa^{-1}$.
  }
    \label{fig:generic}
\end{figure}

Uninhibited infected population $n$ growths in time $t$ according to the Malthus law \cite{Malthus:1798}
\begin{equation}
  \label{eq-malthus}
  \frac{d}{d t} n_{e} = \kappa n_{e} 
\end{equation}
The intrinsic population-independent rate $\kappa$ entering \gl~(\ref{eq-malthus})
is expressed in terms of the probability $\beta$ of infection per encounter with an infected individual
multiplied by the number $\mathcal{N}$ of encounters per unit time (day)
\begin{equation}
  \label{eq-kappa}
  \kappa = \beta \mathcal{N}
\end{equation}
This yields an unlimited exponential time growth ($n_0 \equiv n\left(t=0\right)$)
\begin{equation}
  \label{eq-exp}
  n_{e}(t) = n_0 e^{\kappa t} 
\end{equation}
depicted by the dark green J-shaped curve of \figname\ref{fig:generic}a.

In a real situation, the exponential growth will gradually slow down and eventually level off. 
Infections become more and more unlikely
because, in a given environment, 
the increase in the number of infected diminishes the number of individuals that can be infected.
Rephrasing, the effective growth rate decreases with increasing population density:
$\kappa \to \tilde{\kappa} = f(n) \kappa < \kappa$. 
By assuming 
a linear decrease of $f$ with $n$, one arrives at the logistic model
\cite{Quetelet:1848,Ostwald:1883,McKendrick:1912,Lloyd:67,Cramer:02,Vandermeer:10}
\begin{eqnarray}
  \label{eq-effective-kappa}
  \tilde{\kappa} & & = \kappa \left( 1 - \frac{n}{N}\right) \\
  \label{eq-logistic}
   \frac{d}{d t} n & & \xlongequal{\tilde{\kappa} = \kappa (1 - n/N)} \kappa n \left(1 - \frac{n}{N}\right) 
\end{eqnarray}
Plotted as a function of time (\figname\ref{fig:generic}a), 
\gl~(\ref{eq-logistic}) yields an exponential J-shaped curve only at early times
which switches to an S-shaped (sigmoid) curve as the population increases and saturates to 
the maximum (plateau) value $N$, which defines the so-called carrying capacity of a given environment.
The parenthesis entering the right hand side of \gl~(\ref{eq-logistic}), which
acts as Darwin's ``struggle for existence'' and suppresses 
the exponential growth,
is similar to the Pauli blocking factor extensively discussed in electron transport theory 
\cite{Datta:92,Sols:92,Schoenhammer:93,Datta:97,Wagner:00}.

The continuous time representation underlying the differential 
\gl~(\ref{eq-logistic}) allows to express the
cumulative number of cases $n(t)$ in closed analytical form
\begin{eqnarray}
  \label{eq-f}
  n(t=0) & = & n_0 \\
  n(t) & = & \frac{N}{1 + \left(\frac{N}{n_0} - 1\right) e^{-\kappa t}} 
\end{eqnarray}
By using the half-time $T_{50}$
\begin{equation}
  \label{eq-T50-def}
  n\left(t=T_{50}\right) = \frac{N}{2}
\end{equation}
which defines the 
crossover point where the population attains $p=50$ percent of its maximum
($N$), the above results can be recast as follows
\begin{eqnarray}
  \label{eq-f50}
  n(t) & = & \frac{N}{1 + e^{-\kappa \left( t - T_{50}\right)}}\\
  \label{eq-T50}
  T_{50} & = & \frac{1}{\kappa} \ln\left(\frac{N}{n_0} - 1\right) 
\end{eqnarray}
At $t=T_{50}$, there is a substantial infection slowing with respect to the exponential growth.
There, the instantaneous infection rate $\tilde{\kappa}$ is reduced by 50\% as compared to that 
of the uninhibited growth $\kappa$ (\gl~(\ref{eq-effective-kappa}) and \figname\ref{fig:generic}b).

Saturation occurs within a few characteristic times $T_{c} \equiv \kappa^{-1}$
beyond the half-time $T_{50}$ (\figname\ref{fig:generic}a).
The moment (``day'') $T_{p}$ when
the number $N_{p}$ of infected amounts to $p$ percent
(e.g., $p=95 \mbox{ or } 99$, cf.~\figname\ref{fig:generic})
of the maximum value $N$
\begin{equation}
  \label{eq-Tp-def}
   n\left(t=T_{p}\right) \equiv \frac{p}{100} N
\end{equation}
can easily be deduced from \gl~(\ref{eq-f50})
\begin{eqnarray}
  \label{eq-Tp}
  T_{p} & = & \frac{1}{\kappa} \ln \frac{N/n_0 - 1}{100/p - 1}
  = T_{50} + \frac{1}{\kappa} \ln\frac{p}{100 - p}
\end{eqnarray}
The quantity $\kappa$ is important because it quantifies the daily new infected cases 
expressed by the time derivative $d n/d t$ (\gl~(\ref{eq-df}) and the red curve in \figname\ref{fig:generic}c)
\begin{equation}
  \label{eq-df}
  \frac{d}{d t} n(t) = \frac{N \kappa}{4} \mbox{sech}^2 \frac{\kappa \left( t - T_{50}\right)}{2} \\
\end{equation}
The height $H$ and full width at half maximum FWHM of this so-called
epidemiological curve $d n/d t = f(t)$ are expressed by
\begin{eqnarray}
  \label{eq-H} H & \equiv & {\max}_{t} \frac{d}{d t} n(t) = \frac{1}{4} \kappa N \\
  \label{eq-fwhm}  \mbox{FWHM} & = &
  1.763 / \kappa
\end{eqnarray}
Large values of $\kappa$, amounting to a sharp and high peak in \figname\ref{fig:generic}c, 
may cause healthcare systems collapses.

Diminishing the value of $\kappa$ is of paramount practical importance for a twofold reason:

(i) It renders the peak broader and smaller (\figname\ref{fig:generic}c).
Keeping the daily number of infected cases at a manageable level
(``flattening of the curve'') is essential 
for not overwhelming the healthcare system beyond its capacity to treat the sick.

(ii) Small values of $\kappa$ yield large values of $T_{50}$ (cf.~\gl~(\ref{eq-T50})).
This means postponement of infection explosion and hence gaining time for
a better sanitary and logistic preparation to tackle an upcoming problem:
preventing shortage of intensive care unit beds 
and gaining time for
securing 
and/or producing critical emergency 
  equipment (e.g., masks, ventilators,
  artificial lungs,
  personal protective equipment, 
  extracorporeal membrane oxygenation machines, 
  ventilators or other devices) needed to reduce mortality rate.

\subsection{Modeling Infection in Interacting Groups}
\label{sec:groups}
\begin{figure*}
  \centerline{\includegraphics[width=0.9\textwidth]{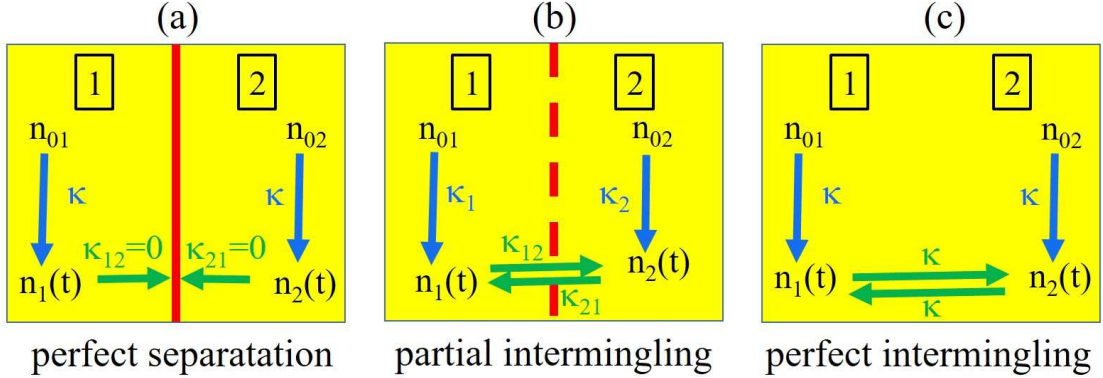}}
  \caption{Schematic representation of a network consisting of two groups 1 and 2
    wherein epidemic can spread (panel b) both through intra-group infections
    (infection rates $\kappa_{1}$ and $\kappa_{2}$) and through
    inter-group infections (infection rates $\kappa_{12}$ and $\kappa_{21}$).
    Limiting cases wherein the network is split into two groups that
    are completely separated among themselves ($\kappa_{12} = \kappa_{21} = 0$, panel a) or
    perfectly intermingled ($\kappa_{1,2} = \kappa_{12} = \kappa_{21} = \kappa$, panel c).
  }
    \label{fig:schemes-covid19}
\end{figure*}

The above considerations referred to a closed population in which members are neither added nor lost from the
group. Neither ``imported'' nor ``exported'' infections were included.
Let us now focus on a network consisting of groups of individuals
$\left\{n_j \right\} = \left\{n_1, n_2, \ldots, n_{m}\right\}$
wherein infections can proliferate both by infections within the same 
group (intra-group infection rates $\kappa_{j} \equiv \kappa_{jj}$)
and because individuals of one group $j$ can infect or can be infected
by individuals of other groups $p\neq j$ (inter-group/intermingling infection rates
$\kappa_{pj}$ and $\kappa_{jp}$, respectively). 
\figname\ref{fig:schemes-covid19} schematically depicts the case of two groups. 

Generalizing the idea underlying \gl~(\ref{eq-logistic}), we will consider below the
following extended logistic model
\begin{eqnarray} 
  \frac{d}{d t} n_{j} & & = \sum_{p=1}^{m} \kappa_{jp} n_{p} \left(1 - \frac{n_{j}}{N_{j}}\right) 
\label{eq-general-coupling}
\end{eqnarray}

Although solving \gl~(\ref{eq-general-coupling}) in general poses no special numerical problem
(some examples are presented in \figname\ref{fig:n-vs-4n} and \figname\ref{fig:n0_0.1_vs-n0_0.001} of
\secname\ref{sec:additional}),
to better emphasize the strategy we aim at conveying, let us first focus on the case of identical groups
\begin{eqnarray}
  & & n_{j}(t=0) = n_0 \equiv n_{0T}/m \label{eq-n0T} \\
  & & N_j = N = N_{T}/m \nonumber \\
  & &\kappa_{jj} = \kappa; \ \kappa_{j\neq p} = \kappa^{\prime}
\label{eq-identical-groups}
\end{eqnarray}
\Gl~(\ref{eq-identical-groups}) yields $j$-independent populations
\begin{equation}
\label{eq-n-total}
  n_{j}(t) = \frac{1}{m} \sum_{j=1}^{m} n_{j}(t) \equiv \frac{1}{m} n_{T}(t)
\end{equation}
$n_{T}(t)$ being the total time dependent population.

Upon term-by-term addition ($\sum_{j=1}^{m} \ldots$) of \gl~(\ref{eq-general-coupling}) we immediately get
\begin{equation}
\label{eq-dn-total}
  \frac{d}{d t}n_{T}(t) = \left[ \kappa + (m - 1)\kappa^{\prime}\right] n_{T}(t) \left(1 - \frac{n_{T}(t)}{N_{T}}\right)
\end{equation}
and hence
\begin{eqnarray}
  \label{eq-f-sym}
  n_{T}(t=0) & = & n_{0T} \\
  n_{T}(t) & = & 
  \frac{N_{T}}{1 +
    \exp\left\{
    -\left[\kappa + (m - 1)\kappa^{\prime}\right]
    \left( t - T_{50}\right) \right\}}  \\
  \label{eq-df-sym}
  \frac{d}{d t} n_{T}(t) & = &
  \frac{\left[\kappa + (m - 1) \kappa^{\prime}\right] N_{T}}{4}
  \mbox{sech}^2 \left[\frac{\kappa + (m - 1) \kappa^{\prime}}{2} \left( t - T_{50}\right)\right] \\
  \label{eq-Tp-sym}
  T_{p} & = & \frac{1}{ \kappa + (m - 1) \kappa^{\prime} } \ln\frac{N_T/n_{0T} - 1}{100/p - 1} \\
  \label{eq-fwhm-sym}
  \mbox{FWHM} & = & \frac{1.763}{\kappa + (m - 1) \kappa^{\prime} }
\end{eqnarray}
By comparing the above formulas with \gls~(\ref{eq-f})-(\ref{eq-T50})
valid for a single group one can conclude that the quantity
\begin{equation}
  \label{eq-kappa-total}
  \kappa_{T} = \kappa + (m - 1) \kappa^{\prime}
\end{equation}
plays the role of a total infection rate.
Importantly, both the half-time $T_{50}$ and the maximum number of daily infections $H$ deduced from
\gls~(\ref{eq-df-sym}) and (\ref{eq-Tp-sym}) 
\begin{eqnarray}
  \label{eq-T50-sym}
  T_{50} & = & \frac{1}{\kappa + (m - 1) \kappa^{\prime}} \ln\left(\frac{N_{T}}{n_{0T}} - 1\right)
  =  \frac{1}{\kappa_T} \ln\left(\frac{N_{T}}{n_{0T}} - 1\right) \\
  \label{eq-H-sym}
  H & \equiv & {\max}_{t} \frac{d}{d t} n_{T}(t) =
  \frac{1}{4}\left[\kappa + (m - 1) \kappa^{\prime} \right] N_T = \frac{1}{4} \kappa_{T} N_{T}
\end{eqnarray}
are controlled by $\kappa_{T}$.
\subsection{Analysis of Two Limiting Cases of Practical Importance}
\label{sec:limiting-cases}
The results of \secname\ref{sec:groups} allow us to
compare how infection propagates in a network (``larger group'') of individuals
split into several ($m$) smaller (sub)groups (chosen identical for simplicity) which do not interact
with each other against the case of a fictitious splitting, wherein (from the point of view of infection)
interactions
between members of a given group and between members belonging to different groups
are identical (perfect intermingling).
Noteworthy, whether the groups are completely separated of each other (label $s$) or
perfectly intermingled (label $i$), the total initial population $n_{0T} = m n_{0}$
is taken to be the same in both cases (cf.~\gl~(\ref{eq-n0T}))
\begin{equation}
  \label{eq-same-n0T}
  n_{0T}^{s} = n_{0T}^{i}
\end{equation}

The former case, corresponding to separated groups (label $i$),
is characterized by a vanishing inter-group infection rate ( $\kappa^{\prime} \equiv 0 $). 
Applied to this case, \gls~(\ref{eq-kappa-total}), (\ref{eq-T50-sym}) and (\ref{eq-H-sym}) yield 
\begin{equation}
  \label{eq-isolated}
  \kappa_{T}^{s} = \kappa;\ T_{50}^{s} = \frac{1}{\kappa}\ln \left(\frac{N_{T}}{n_{T0}} - 1\right);\ H^{s} = \frac{1}{4}\kappa N_{T}
\end{equation}

At the opposite extreme of perfectly intermingled groups,
inter-group interactions are as strong as intra-group interactions
($\kappa^{\prime} = \kappa$). Based on \gls~(\ref{eq-kappa-total}), (\ref{eq-T50-sym}) and (\ref{eq-H-sym})
we then get
\begin{equation}
  \label{eq-permeated}
  \kappa_{T}^{i} = m \kappa;\ T_{50}^{i} = \frac{1}{m \kappa}\ln \left(\frac{N_{T}}{n_{T0}} - 1\right);\ H^{i} = \frac{1}{4} m \kappa N_{T}
\end{equation}

The above results show that the epidemiological curve can be substantially flattened if a
larger group is split into several smaller groups separated from each other.
By starting from the same number of infections (\gl~(\ref{eq-same-n0T})),
splitting into groups separated 
from one another yields a reduction of the infection rate and of the maximum daily cases by a factor $m$
and an increase of full width at half maximum by the same factor
\begin{eqnarray}
  \label{eq-kappa-i-vs-p}
  \kappa_{T}^{s} & = & \frac{\kappa^{s}}{m} \\
  \label{eq-fwhm-i-vs-p}
  \mbox{FHWM}^{s} & = & m\, \mbox{FWHM}^{i} \\
  \label{eq-H-i-vs-p}
  H^{s} & = & \frac{H^{i}}{m} 
\end{eqnarray}

Equally pleasantly, splitting leads in addition
to a time postponement of the infection peak by the same factor $m$
\begin{equation} 
  \label{eq-T50-i-vs-p} 
  T_{50}^{s} = m\, T_{50}^{i}
\end{equation}

\begin{figure}
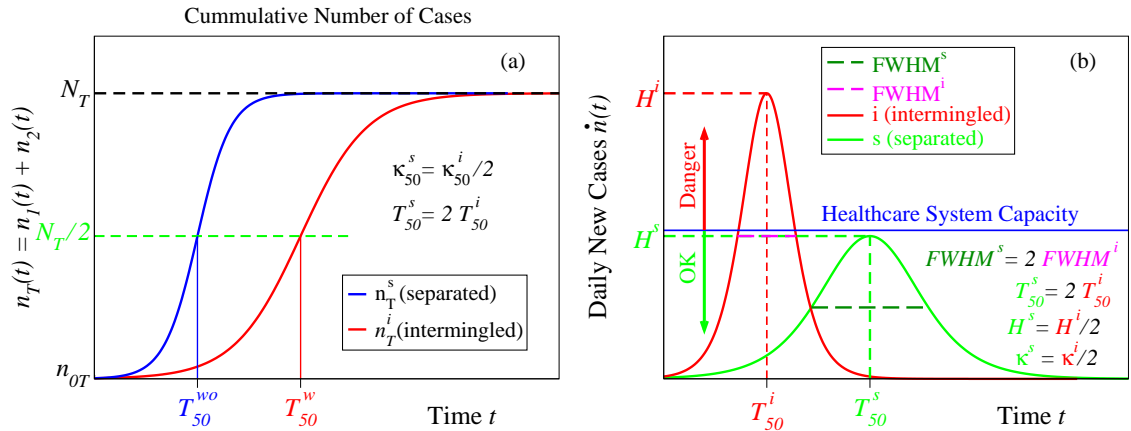

  \centerline{\includegraphics[width=0.45\textwidth]{fig_f_k_2_vs_k_4}
    \includegraphics[width=0.45\textwidth]{fig_df_k_2_vs_k_4}
  }
  \caption{Results demonstrating how preventing intermingling between several groups flattens the epidemic curve
    with the concomitant time delay of the maximum of daily infections.
    (a) Cumulative number of cases $n_{1,2}(t)$ and (b)
    epidemiological curve $\dot{n}_{1,2}(t) \equiv d n_{1,2}(t)$
    for a group consisting
    of two subgroups that are either completely separated of each other (green lines)
    or perfectly intermingled (red curves). These results also schematically depict 
    how flattening and delaying the epidemiological curve by a factor $m=2$ by
    splitting into $m=2$ groups can prevent overloading the healthcare system capacity.
    Time on the $x$-axis is expressed in units of the characteristic time
    $T_{c} =\kappa^{-1}$.}
    \label{fig:c-vs-c}
\end{figure}

The results presented in \figname\ref{fig:c-vs-c} depict these findings for the particular case
of a larger group split into two smaller groups ($m=2$). They also schematically visualize how
and why group splitting can relieve the healthcare system.
To avoid misunderstandings, one should note that the value $m=2$ in \figname\ref{fig:c-vs-c} was chosen
just for more clarity.  
Splitting into more than two noninteracting groups (i.e., making $m$ as large as possible) 
is highly desirable.

Before ending this part, we want to emphasize that, however important,
flattening and delaying the epidemiological curve
by a factor $m$ achieved by group splitting (cf.~\figname\ref{fig:c-vs-c} and
\gls~(\ref{eq-kappa-i-vs-p}), (\ref{eq-fwhm-i-vs-p}), and (\ref{eq-H-i-vs-p}))
is not the whole issue. Extremely importantly, the presently
proposed group splitting approach does not assume any intra-group 
(like social distancing and wearing masks) restrictions: economy and social life within individual 
groups separated of each other can continue.
\subsection{Additional Results}
\label{sec:additional}
\begin{figure}
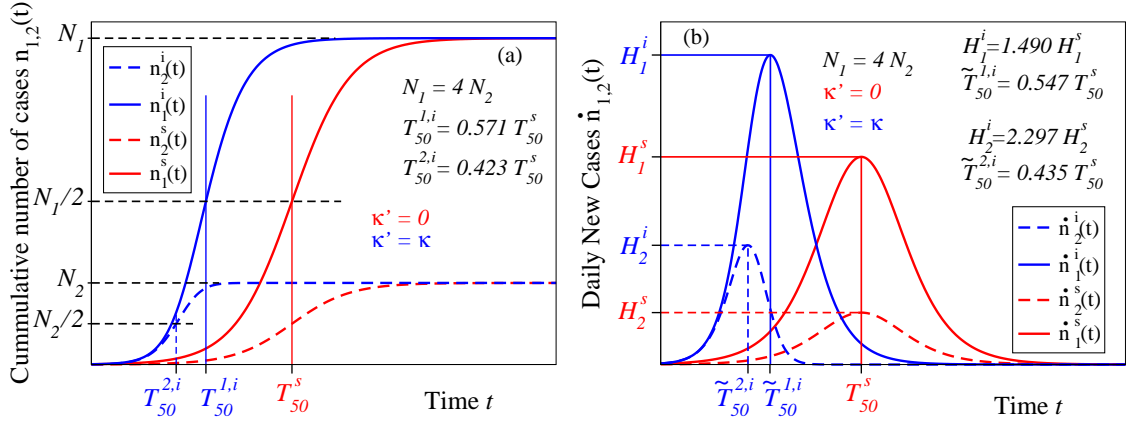

  \centerline{\includegraphics[width=0.45\textwidth]{fig_f_k_2_n_vs_4n}
    \includegraphics[width=0.45\textwidth]{fig_df_k_2_n_vs_4n}
  }
  \caption{(a) Cumulative number of cases $n_{1,2}(t)$ and (b) epidemiological curves
    $\dot{n}_{1,2}(t) \equiv d n_{1,2}(t)/d t$ for two
    completely separated and perfectly intermingled (red and blue curves, respectively)
    groups whose populations differ by a factor of four.
    Notice that, in spite of the equal infection rates, infection of the
    smaller group 2 (dashed lines) is stronger affected by the inter-group interaction 
    than in the larger group 1 (solid lines).
    Time on the $x$-axis is expressed in units of the characteristic time
    $T_{c} =\kappa^{-1}$.}
  \label{fig:n-vs-4n} 
\end{figure}

\begin{figure}
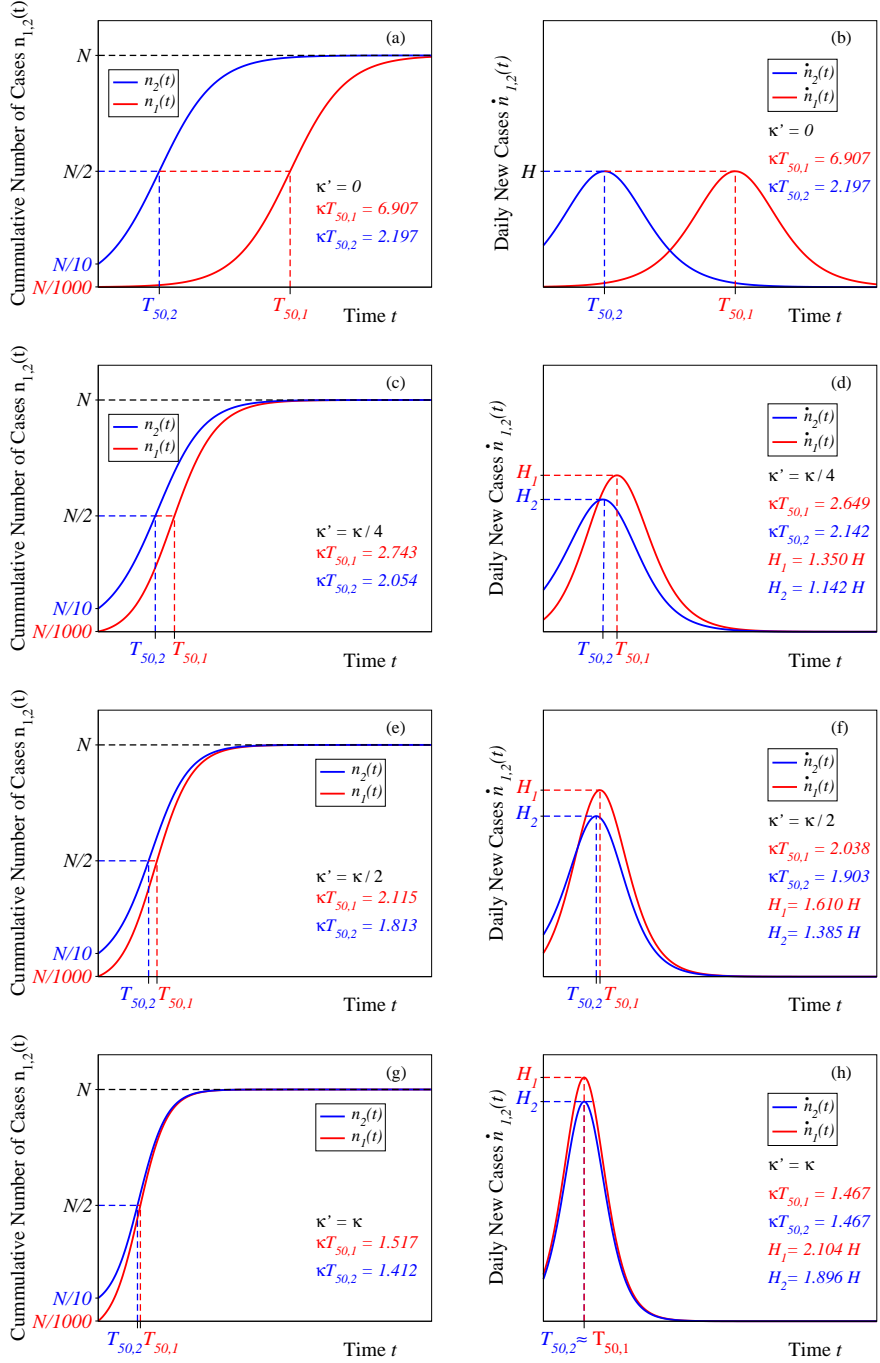

  \centerline{\includegraphics[width=0.35\textwidth]{fig_f_kd_2_kn_0_n_1_n01_0.001_n02_0.1.eps}
    \includegraphics[width=0.35\textwidth]{fig_df_kd_2_kn_0_n_1_n01_0.001_n02_0.1.eps}
  }
  \centerline{\includegraphics[width=0.35\textwidth]{fig_f_kd_2_kn_0.5_n_1_n01_0.001_n02_0.1.eps}
    \includegraphics[width=0.35\textwidth]{fig_df_kd_2_kn_0.5_n_1_n01_0.001_n02_0.1.eps}
  }
  \centerline{\includegraphics[width=0.35\textwidth]{fig_f_kd_2_kn_1_n_1_n01_0.001_n02_0.1.eps}
    \includegraphics[width=0.35\textwidth]{fig_df_kd_2_kn_1_n_1_n01_0.001_n02_0.1.eps}
  }
  \centerline{\includegraphics[width=0.35\textwidth]{fig_f_kd_2_kn_2_n_1_n01_0.001_n02_0.1.eps}
    \includegraphics[width=0.35\textwidth]{fig_df_kd_2_kn_2_n_1_n01_0.001_n02_0.1.eps}
  }
  \caption{Results showing the impact of the inter-group interaction $\kappa^{\prime}$
    on the cumulative number of cases $n_{1,2}(t)$ and
    epidemiological curves $\dot{n}_{1,2}(t) \equiv d n_{1,2}(t) / d t $ in case of two groups merely differing
    by the initial number of infected cases: $n_{01}/N = 0.001$ and $n_{02}/N = 0.1$.
    They show how infections in the initially less infected group 1 are rapidly triggered by
    infections in the initially more infected group 2.
    Time on the $x$-axis is expressed in units of the characteristic time
    $T_{c} =\kappa^{-1}$.}
  \label{fig:n0_0.1_vs-n0_0.001} 
\end{figure}

As anticipated in \secname\ref{sec:groups}, the logistic model
extended as expressed by \gl~(\ref{eq-general-coupling}) can be used to quantify the
impact of mutual infections in networks of interacting groups
more general than the particular situations examined in \secname\ref{sec:limiting-cases}.
To briefly illustrate this fact, two examples are presented in \figsname\ref{fig:n-vs-4n}
and \ref{fig:n0_0.1_vs-n0_0.001}.

\figname\ref{fig:n-vs-4n} depicts the case two groups whose populations differ by a factor of four.
As visible there, in spite of the equal infection rates, infection of the
smaller group 2 (dashed lines) is stronger enhanced by the inter-group interaction 
than in the larger group 1 (solid lines).

\figname\ref{fig:n0_0.1_vs-n0_0.001} presents the case two groups merely differing from each other
by the different numbers of initially infected individuals ($n_{01}/N = 0.001$ versus $n_{02}/N = 0.1$).
Comparison of the various panels of \figname\ref{fig:n0_0.1_vs-n0_0.001} reveals that
inter-group infection yields an infection rapidly ``exported'' from the
initially more infected group to that which was initially less infected.  
Intermingling (\figname\ref{fig:n0_0.1_vs-n0_0.001}g) quickly wipes out
any difference between a initially weaker (or non)infected group
and an initially strongly infected group.
\section{Conclusion}
\label{sec:conclusion}
In closing, the results presented above indicate that
splitting of a large group into smaller groups and reducing intermingling 
appears to be an appealing strategy for substantially reducing the spread of the SARS-CoV-2 virus
while still allowing 
social life to a more bearable level than distancing everyone from anyone
and economy go on, albeit slower.

One can expect that (A) it is easier to impose and maintain longer-term
regulations on splitting a given population into
(say, 5 -- 10) groups weakly interacting among themselves
than (B) enforcing severe containment measures to all individuals
in order to diminish the probability $\beta$ of infection per encounter
and the number $\mathcal{N}$ of daily encounters
yielding a reduction by the same factor (5--10) of the infection rate $\kappa$
(cf.~\gl~(\ref{eq-kappa})). The big difference is that with option (A) 
social life and economy within individual groups go on without
intra-group restrictions, while option (B) means attaining the \emph{same} epidemiological curve
with both social life and economy paralyzed. 

Devising and implementing an adequate restructuring of large logistic units
(large productive networks, factories, enterprises, warehouses, etc) allowing, if/when necessary,
society to rapidly switch back and forth between 
separated groups and intermingled groups can certainly be challenging
but may be a long-term strategic goal worth to be pursued when faced with these
COVID-19 pandemic times or other similar difficulties that cannot be ruled out in the future.
Last but not least, the declaration on 15 May 2020, for the first time in Europe, that COVID-19 epidemic ended in
the two million people Slovenia \cite{EndCOVID19-Slovenia,Baldea:2020f}
may be taken as confirming the advantage of the strategy
proposed in this paper.
%
%
%
%
%
%
%
%

%
%
%
%

\begin{mcitethebibliography}{42}
\providecommand*\natexlab[1]{#1}
\providecommand*\mciteSetBstSublistMode[1]{}
\providecommand*\mciteSetBstMaxWidthForm[2]{}
\providecommand*\mciteBstWouldAddEndPuncttrue
  {\def\EndOfBibitem{\unskip.}}
\providecommand*\mciteBstWouldAddEndPunctfalse
  {\let\EndOfBibitem\relax}
\providecommand*\mciteSetBstMidEndSepPunct[3]{}
\providecommand*\mciteSetBstSublistLabelBeginEnd[3]{}
\providecommand*\EndOfBibitem{}
\mciteSetBstSublistMode{f}
\mciteSetBstMaxWidthForm{subitem}{(\alph{mcitesubitemcount})}
\mciteSetBstSublistLabelBeginEnd
  {\mcitemaxwidthsubitemform\space}
  {\relax}
  {\relax}

\bibitem[Cucinotta and Vanelli(2020)Cucinotta, and
  Vanelli]{WHO_COVID-19_Pandemic}
Cucinotta,~D.; Vanelli,~M. WHO Declares COVID-19 a Pandemic. \emph{Acta Bio
  Medica Atenei Parmensis} \textbf{2020}, \emph{91}, 157--160\relax
\mciteBstWouldAddEndPuncttrue
\mciteSetBstMidEndSepPunct{\mcitedefaultmidpunct}
{\mcitedefaultendpunct}{\mcitedefaultseppunct}\relax
\EndOfBibitem
\bibitem[Baldwin and di~Mauro(2020)Baldwin, and di~Mauro]{Baldwin:20}
Baldwin,~R., di~Mauro,~B.~W., Eds. \emph{Mitigating the COVID Economic Crisis:
  Act Fast and Do Whatever It Takes}; A CEPR Press VoxEU.org eBook, 2020\relax
\mciteBstWouldAddEndPuncttrue
\mciteSetBstMidEndSepPunct{\mcitedefaultmidpunct}
{\mcitedefaultendpunct}{\mcitedefaultseppunct}\relax
\EndOfBibitem
\bibitem[Chen \latin{et~al.}(2020)Chen, Rui, Wang, Zhao, Cui, and
  Yin]{Chen2020.01.19.911669}
Chen,~T.; Rui,~J.; Wang,~Q.; Zhao,~Z.; Cui,~J.-A.; Yin,~L. A mathematical model
  for simulating the transmission of Wuhan novel Coronavirus. \emph{bioRxiv}
  \textbf{2020}, DOI 10.1101/2020.01.19.911669\relax
\mciteBstWouldAddEndPuncttrue
\mciteSetBstMidEndSepPunct{\mcitedefaultmidpunct}
{\mcitedefaultendpunct}{\mcitedefaultseppunct}\relax
\EndOfBibitem
\bibitem[Su \latin{et~al.}(2020)Su, Hong, He, Ma, Wang, Xu, Jiang, Han, Chang,
  Zhu, and Long]{Su2020.03.06.20032177}
Su,~L.; Hong,~N.; He,~J.; Ma,~Y.; Wang,~X.; Xu,~Q.; Jiang,~H.; Han,~L.;
  Chang,~F.; Zhu,~W.; Long,~Y. Evaluation of the secondary transmission pattern
  and epidemic prediction of COVID-19 in the four metropolitan areas of China.
  \emph{medRxiv} \textbf{2020}, DOI 10.1101/2020.03.06.20032177\relax
\mciteBstWouldAddEndPuncttrue
\mciteSetBstMidEndSepPunct{\mcitedefaultmidpunct}
{\mcitedefaultendpunct}{\mcitedefaultseppunct}\relax
\EndOfBibitem
\bibitem[Q \latin{et~al.}(2020)Q, X, and et~al.]{Li:2020}
Q,~L.; X,~G.; et~al.,~W.~P. Early Transmission Dynamics in Wuhan, China, of
  Novel Coronavirus-Infected Pneumonia. \emph{N Engl J Med.} \textbf{2020},
  \emph{382}, 1199--1207\relax
\mciteBstWouldAddEndPuncttrue
\mciteSetBstMidEndSepPunct{\mcitedefaultmidpunct}
{\mcitedefaultendpunct}{\mcitedefaultseppunct}\relax
\EndOfBibitem
\bibitem[Roosa \latin{et~al.}(2020)Roosa, Lee, and Luo]{Roosa:2020}
Roosa,~K.; Lee,~Y.; Luo,~R. e.~a. Short-term Forecasts of the COVID-19 Epidemic
  in Guangdong and Zhejiang, China: February 13-23, 2020. \emph{J. Clin. Med.}
  \textbf{2020}, \emph{9}, 596\relax
\mciteBstWouldAddEndPuncttrue
\mciteSetBstMidEndSepPunct{\mcitedefaultmidpunct}
{\mcitedefaultendpunct}{\mcitedefaultseppunct}\relax
\EndOfBibitem
\bibitem[Shen \latin{et~al.}(2020)Shen, Peng, Xiao, and
  Zhang]{Shen2020.01.23.916726}
Shen,~M.; Peng,~Z.; Xiao,~Y.; Zhang,~L. Modelling the epidemic trend of the
  2019 novel coronavirus outbreak in China. \emph{bioRxiv} \textbf{2020}, DOI
  10.1101/2020.01.23.916726\relax
\mciteBstWouldAddEndPuncttrue
\mciteSetBstMidEndSepPunct{\mcitedefaultmidpunct}
{\mcitedefaultendpunct}{\mcitedefaultseppunct}\relax
\EndOfBibitem
\bibitem[Hui \latin{et~al.}(2020)Hui, {I Azhar}, Madani, Ntoumi, Kock, Dar,
  Ippolito, Mchugh, Memish, Drosten, Zumla, and
  Petersen]{8b30d75ac97f46a0bafcad5166faa39f}
Hui,~D.; {I Azhar},~E.; Madani,~T.; Ntoumi,~F.; Kock,~R.; Dar,~O.;
  Ippolito,~G.; Mchugh,~T.; Memish,~Z.; Drosten,~C.; Zumla,~A.; Petersen,~E.
  The continuing 2019-nCoV epidemic threat of novel coronaviruses to global
  health — The latest 2019 novel coronavirus outbreak in Wuhan, China.
  \emph{International Journal of Infectious Diseases} \textbf{2020}, \emph{91},
  264--266\relax
\mciteBstWouldAddEndPuncttrue
\mciteSetBstMidEndSepPunct{\mcitedefaultmidpunct}
{\mcitedefaultendpunct}{\mcitedefaultseppunct}\relax
\EndOfBibitem
\bibitem[Yang \latin{et~al.}(2020)Yang, Qi, Zhang, Wang, Bi, Yang, and
  Sheng]{Yang2020.04.01.20043794}
Yang,~P.; Qi,~J.; Zhang,~S.; Wang,~X.; Bi,~G.; Yang,~Y.; Sheng,~B. Feasibility
  Study of Mitigation and Suppression Intervention Strategies for Controlling
  COVID-19 Outbreaks in London and Wuhan. \emph{medRxiv} \textbf{2020}, DOI
  10.1101/2020.04.01.20043794\relax
\mciteBstWouldAddEndPuncttrue
\mciteSetBstMidEndSepPunct{\mcitedefaultmidpunct}
{\mcitedefaultendpunct}{\mcitedefaultseppunct}\relax
\EndOfBibitem
\bibitem[Aggarwal(2020)]{aggarwal2020.04.17.20069245}
Aggarwal,~N. Importance of Social Distancing: Modeling the spread of 2019-nCoV
  using Susceptible-Infected-Quarantined-Recovered-t model. \emph{medRxiv}
  \textbf{2020}, 10.1101/2020.04.17.20069245\relax
\mciteBstWouldAddEndPuncttrue
\mciteSetBstMidEndSepPunct{\mcitedefaultmidpunct}
{\mcitedefaultendpunct}{\mcitedefaultseppunct}\relax
\EndOfBibitem
\bibitem[Kermack and McKendrick(1927)Kermack, and McKendrick]{Kermack:27}
Kermack,~W.~O.; McKendrick,~A.~G. Contributions to the mathematical theory of
  epidemics. I. \emph{Proc. Roy. Soc.} \textbf{1927}, \emph{115A}, 700--–721,
  reprinted in Bull. Math. Biol. 53, 33–55 (1991).
  https://doi.org/10.1007/BF02464423\relax
\mciteBstWouldAddEndPuncttrue
\mciteSetBstMidEndSepPunct{\mcitedefaultmidpunct}
{\mcitedefaultendpunct}{\mcitedefaultseppunct}\relax
\EndOfBibitem
\bibitem[Kermack and McKendrick(1932)Kermack, and McKendrick]{Kermack:32}
Kermack,~W.~O.; McKendrick,~A.~G. Contributions to the mathematical theory of
  epidemics—II. The problem of endemicity. \emph{Proc. Roy. Soc.}
  \textbf{1932}, \emph{138A}, 55--83, reprinted in Bull. Math. Biol. 53,
  57–87 (1991). https://doi.org/10.1007/BF02464424\relax
\mciteBstWouldAddEndPuncttrue
\mciteSetBstMidEndSepPunct{\mcitedefaultmidpunct}
{\mcitedefaultendpunct}{\mcitedefaultseppunct}\relax
\EndOfBibitem
\bibitem[Kermack and McKendrick(1933)Kermack, and McKendrick]{Kermack:33}
Kermack,~W.~O.; McKendrick,~A.~G. Contributions to the mathematical theory of
  epidemics. III. Further studies of the problem of endemicity. \emph{Proc.
  Roy. Soc.} \textbf{1933}, \emph{141A}, 94--122, reprinted in Bull. Math.
  Biol. 53, 89–118 (1991). https://doi.org/10.1007/BF02464425\relax
\mciteBstWouldAddEndPuncttrue
\mciteSetBstMidEndSepPunct{\mcitedefaultmidpunct}
{\mcitedefaultendpunct}{\mcitedefaultseppunct}\relax
\EndOfBibitem
\bibitem[Bailey(1975)]{Bailey:75}
Bailey,~N. T.~J. \emph{The Mathematical Theory of Infectious Diseases and Its
  Applications}; Charles Griffin \& Company Ltd, 5a Crendon Street, High
  Wycombe, Bucks HP13 6LE., 1975\relax
\mciteBstWouldAddEndPuncttrue
\mciteSetBstMidEndSepPunct{\mcitedefaultmidpunct}
{\mcitedefaultendpunct}{\mcitedefaultseppunct}\relax
\EndOfBibitem
\bibitem[Hethcote(1994)]{Hethcote:94}
Hethcote,~H.~W. A Thousand and One Epidemic Models. Frontiers in Mathematical
  Biology. Berlin, Heidelberg, 1994; pp 504--515\relax
\mciteBstWouldAddEndPuncttrue
\mciteSetBstMidEndSepPunct{\mcitedefaultmidpunct}
{\mcitedefaultendpunct}{\mcitedefaultseppunct}\relax
\EndOfBibitem
\bibitem[Hethcote(2000)]{Hethcote:00}
Hethcote,~H.~W. The Mathematics of Infectious Diseases. \emph{SIAM Review}
  \textbf{2000}, \emph{42}, 599--653\relax
\mciteBstWouldAddEndPuncttrue
\mciteSetBstMidEndSepPunct{\mcitedefaultmidpunct}
{\mcitedefaultendpunct}{\mcitedefaultseppunct}\relax
\EndOfBibitem
\bibitem[Morris(2020)]{SEIR_simulator}
Morris,~S.~E. shinySIR: Interactive Plotting for Mathematical Models of
  Infectious Disease Spread. 2020; R package version 0.1.1\relax
\mciteBstWouldAddEndPuncttrue
\mciteSetBstMidEndSepPunct{\mcitedefaultmidpunct}
{\mcitedefaultendpunct}{\mcitedefaultseppunct}\relax
\EndOfBibitem
\bibitem[Mat()]{MathematicaSEIR}
Epidemiological Models for Influenza and COVID-19.
  \url{https://community.wolfram.com/groups/-/m/t/1896178},
  https://community.wolfram.com/groups/-/m/t/1896178\relax
\mciteBstWouldAddEndPuncttrue
\mciteSetBstMidEndSepPunct{\mcitedefaultmidpunct}
{\mcitedefaultendpunct}{\mcitedefaultseppunct}\relax
\EndOfBibitem
\bibitem[Goh(2020)]{EpidemicCalculator}
Goh,~G. Epidemic Calculator. 2020;
  \url{http://gabgoh.github.io/COVID/index.html},
  http://gabgoh.github.io/COVID/index.html\relax
\mciteBstWouldAddEndPuncttrue
\mciteSetBstMidEndSepPunct{\mcitedefaultmidpunct}
{\mcitedefaultendpunct}{\mcitedefaultseppunct}\relax
\EndOfBibitem
\bibitem[Fong \latin{et~al.}(2020)Fong, Li, Dey, Crespo, and
  Herrera-Viedma]{IJIMAI-3841}
Fong,~S.~J.; Li,~G.; Dey,~N.; Crespo,~R.~G.; Herrera-Viedma,~E. Finding an
  Accurate Early Forecasting Model from Small Dataset: A Case of 2019-nCoV
  Novel Coronavirus Outbreak. \emph{International Journal of Interactive
  Multimedia and Artificial Intelligence} \textbf{2020}, \emph{6},
  132--140\relax
\mciteBstWouldAddEndPuncttrue
\mciteSetBstMidEndSepPunct{\mcitedefaultmidpunct}
{\mcitedefaultendpunct}{\mcitedefaultseppunct}\relax
\EndOfBibitem
\bibitem[Quetelet(1848)]{Quetelet:1848}
Quetelet,~L. A.~J. \emph{Du Syst{\`e}me Social et des Lois qui le
  R{\'e}gissent}; Guillaumin, 1848\relax
\mciteBstWouldAddEndPuncttrue
\mciteSetBstMidEndSepPunct{\mcitedefaultmidpunct}
{\mcitedefaultendpunct}{\mcitedefaultseppunct}\relax
\EndOfBibitem
\bibitem[Ostwald(1883)]{Ostwald:1883}
Ostwald,~W. Studien zur chemischen Dynamik; Erste Abhandlung: Die Einwirkung
  der Säuren auf Acetamid. \emph{Journal f\"ur Praktische Chemie}
  \textbf{1883}, \emph{27}, 1--39\relax
\mciteBstWouldAddEndPuncttrue
\mciteSetBstMidEndSepPunct{\mcitedefaultmidpunct}
{\mcitedefaultendpunct}{\mcitedefaultseppunct}\relax
\EndOfBibitem
\bibitem[McKendrick and Pai(1912)McKendrick, and Pai]{McKendrick:1912}
McKendrick,~A.~G.; Pai,~M.~K. XLV. The Rate of Multiplication of
  Micro-organisms: A Mathematical Study. \emph{Proceedings of the Royal Society
  of Edinburgh} \textbf{1912}, \emph{31}, 649--653\relax
\mciteBstWouldAddEndPuncttrue
\mciteSetBstMidEndSepPunct{\mcitedefaultmidpunct}
{\mcitedefaultendpunct}{\mcitedefaultseppunct}\relax
\EndOfBibitem
\bibitem[Lloyd(1967)]{Lloyd:67}
Lloyd,~P. American, German and British antecedents to Pearl and Reed's logistic
  curve. \emph{Population Studies} \textbf{1967}, \emph{21}, 99--108\relax
\mciteBstWouldAddEndPuncttrue
\mciteSetBstMidEndSepPunct{\mcitedefaultmidpunct}
{\mcitedefaultendpunct}{\mcitedefaultseppunct}\relax
\EndOfBibitem
\bibitem[Cramer(2004)]{Cramer:02}
Cramer,~J. The early origins of the logit model. \emph{Studies in History and
  Philosophy of Science Part C: Studies in History and Philosophy of Biological
  and Biomedical Sciences} \textbf{2004}, \emph{35}, 613 -- 626\relax
\mciteBstWouldAddEndPuncttrue
\mciteSetBstMidEndSepPunct{\mcitedefaultmidpunct}
{\mcitedefaultendpunct}{\mcitedefaultseppunct}\relax
\EndOfBibitem
\bibitem[Vandermeer(2010)]{Vandermeer:10}
Vandermeer,~J. How Populations Grow: The Exponential and Logistic Equations.
  \emph{Nature Education Knowledge} \textbf{2010}, \emph{3}, 15\relax
\mciteBstWouldAddEndPuncttrue
\mciteSetBstMidEndSepPunct{\mcitedefaultmidpunct}
{\mcitedefaultendpunct}{\mcitedefaultseppunct}\relax
\EndOfBibitem
\bibitem[B\^aldea(2017)]{Baldea:2017m}
B\^aldea,~I. Floppy Molecules as Candidates for Achieving Optoelectronic
  Molecular Devices without Skeletal Rearrangement or Bond Breaking.
  \emph{Phys. Chem. Chem. Phys.} \textbf{2017}, \emph{19}, 30842 -- 30851\relax
\mciteBstWouldAddEndPuncttrue
\mciteSetBstMidEndSepPunct{\mcitedefaultmidpunct}
{\mcitedefaultendpunct}{\mcitedefaultseppunct}\relax
\EndOfBibitem
\bibitem[B\^aldea(2018)]{Baldea:2018e}
B\^aldea,~I. A sui generis electrode-driven spatial confinement effect
  responsible for strong twisting enhancement of floppy molecules in closely
  packed self-assembled monolayers. \emph{Phys. Chem. Chem. Phys.}
  \textbf{2018}, \emph{20}, 23492--23499\relax
\mciteBstWouldAddEndPuncttrue
\mciteSetBstMidEndSepPunct{\mcitedefaultmidpunct}
{\mcitedefaultendpunct}{\mcitedefaultseppunct}\relax
\EndOfBibitem
\bibitem[Mansfield and Hensley(1960)Mansfield, and Hensley]{Mansfield:60}
Mansfield,~E.; Hensley,~C. The Logistic Process: Tables of the Stochastic
  Epidemic Curve and Applications. \emph{Journal of the Royal Statistical
  Society: Series B (Methodological)} \textbf{1960}, \emph{22}, 332--337\relax
\mciteBstWouldAddEndPuncttrue
\mciteSetBstMidEndSepPunct{\mcitedefaultmidpunct}
{\mcitedefaultendpunct}{\mcitedefaultseppunct}\relax
\EndOfBibitem
\bibitem[Waggoner and Aylor(2000)Waggoner, and Aylor]{Waggoner:00}
Waggoner,~P.~E.; Aylor,~D.~E. Epidemiology: A Science of Patterns. \emph{Annual
  Review of Phytopathology} \textbf{2000}, \emph{38}, 71--94, PMID:
  11701837\relax
\mciteBstWouldAddEndPuncttrue
\mciteSetBstMidEndSepPunct{\mcitedefaultmidpunct}
{\mcitedefaultendpunct}{\mcitedefaultseppunct}\relax
\EndOfBibitem
\bibitem[Koopman(2004)]{Koopman:04}
Koopman,~J. Modeling Infection Transmission. \emph{Annual Review of Public
  Health} \textbf{2004}, \emph{25}, 303--326, PMID: 15015922\relax
\mciteBstWouldAddEndPuncttrue
\mciteSetBstMidEndSepPunct{\mcitedefaultmidpunct}
{\mcitedefaultendpunct}{\mcitedefaultseppunct}\relax
\EndOfBibitem
\bibitem[Bangert \latin{et~al.}(2017)Bangert, Molyneux, Lindsay, Fitzpatrick,
  and Engels]{Bangert:17}
Bangert,~M.; Molyneux,~D.~H.; Lindsay,~S.~W.; Fitzpatrick,~C.; Engels,~D. The
  cross-cutting contribution of the end of neglected tropical diseases to the
  sustainable development goals. \emph{Infect Dis. Poverty} \textbf{2017},
  \emph{6}, 73\relax
\mciteBstWouldAddEndPuncttrue
\mciteSetBstMidEndSepPunct{\mcitedefaultmidpunct}
{\mcitedefaultendpunct}{\mcitedefaultseppunct}\relax
\EndOfBibitem
\bibitem[Hermanowicz(2020)]{Hermanowicz2020.03.31.20049486}
Hermanowicz,~S.~W. Simple Model for Covid-19 Epidemics - Back-casting in China
  and Forecasting in the US. \emph{medRxiv} \textbf{2020}, DOI
  10.1101/2020.03.31.20049486\relax
\mciteBstWouldAddEndPuncttrue
\mciteSetBstMidEndSepPunct{\mcitedefaultmidpunct}
{\mcitedefaultendpunct}{\mcitedefaultseppunct}\relax
\EndOfBibitem
\bibitem[Malthus(1798)]{Malthus:1798}
Malthus,~T.~R. \emph{An Essay on the Principle of Population, as It Affects the
  Future Improvement of Society with Remarks on the Speculations of Mr. Godwin,
  M. Condorcet, and Other Writers}; Printed for J. Johnson, in St. Paul's
  Church-Yard, London, 1798\relax
\mciteBstWouldAddEndPuncttrue
\mciteSetBstMidEndSepPunct{\mcitedefaultmidpunct}
{\mcitedefaultendpunct}{\mcitedefaultseppunct}\relax
\EndOfBibitem
\bibitem[Datta(1992)]{Datta:92}
Datta,~S. Exclusion principle and the Landauer-B\"uttiker formalism.
  \emph{Phys. Rev. B} \textbf{1992}, \emph{45}, 1347--1362\relax
\mciteBstWouldAddEndPuncttrue
\mciteSetBstMidEndSepPunct{\mcitedefaultmidpunct}
{\mcitedefaultendpunct}{\mcitedefaultseppunct}\relax
\EndOfBibitem
\bibitem[Sols(1992)]{Sols:92}
Sols,~F. Scattering, dissipation, and transport in mesoscopic systems.
  \emph{Ann. Phys. (NY)} \textbf{1992}, \emph{214}, 386 -- 438\relax
\mciteBstWouldAddEndPuncttrue
\mciteSetBstMidEndSepPunct{\mcitedefaultmidpunct}
{\mcitedefaultendpunct}{\mcitedefaultseppunct}\relax
\EndOfBibitem
\bibitem[B\"onig and Sch\"onhammer(1993)B\"onig, and
  Sch\"onhammer]{Schoenhammer:93}
B\"onig,~L.; Sch\"onhammer,~K. Pauli principle in the theory of nonlinear
  electronic transport. \emph{Phys. Rev. B} \textbf{1993}, \emph{47},
  9203--9207\relax
\mciteBstWouldAddEndPuncttrue
\mciteSetBstMidEndSepPunct{\mcitedefaultmidpunct}
{\mcitedefaultendpunct}{\mcitedefaultseppunct}\relax
\EndOfBibitem
\bibitem[Datta(1997)]{Datta:97}
Datta,~S. \emph{Electronic Transport in Mesoscopic Systems}; Cambridge Univ.
  Press: Cambridge, 1997\relax
\mciteBstWouldAddEndPuncttrue
\mciteSetBstMidEndSepPunct{\mcitedefaultmidpunct}
{\mcitedefaultendpunct}{\mcitedefaultseppunct}\relax
\EndOfBibitem
\bibitem[Wagner(2000)]{Wagner:00}
Wagner,~M. Probing Pauli Blocking Factors in Quantum Pumps with Broken
  Time-Reversal Symmetry. \emph{Phys. Rev. Lett.} \textbf{2000}, \emph{85},
  174--177\relax
\mciteBstWouldAddEndPuncttrue
\mciteSetBstMidEndSepPunct{\mcitedefaultmidpunct}
{\mcitedefaultendpunct}{\mcitedefaultseppunct}\relax
\EndOfBibitem
\bibitem[End()]{EndCOVID19-Slovenia}
https://www.independent.co.uk/news/world/europe/oronavirus-slovenia-end-epidemic-lockdown-lifted-a9516841.html\relax
\mciteBstWouldAddEndPuncttrue
\mciteSetBstMidEndSepPunct{\mcitedefaultmidpunct}
{\mcitedefaultendpunct}{\mcitedefaultseppunct}\relax
\EndOfBibitem
\bibitem[B\^aldea(2020)]{Baldea:2020f}
B\^aldea,~I. What Can We Learn from the Time Evolution of COVID-19 Epidemic in
  Slovenia? \emph{medRxiv} \textbf{2020}, DOI 10.1101/2020.05.25.20112938\relax
\mciteBstWouldAddEndPuncttrue
\mciteSetBstMidEndSepPunct{\mcitedefaultmidpunct}
{\mcitedefaultendpunct}{\mcitedefaultseppunct}\relax
\EndOfBibitem
\end{mcitethebibliography}
\providecommand{\latin}[1]{#1}
\makeatletter
\providecommand{\doi}
  {\begingroup\let\do\@makeother\dospecials
  \catcode`\{=1 \catcode`\}=2 \doi@aux}
\providecommand{\doi@aux}[1]{\endgroup\texttt{#1}}
\makeatother
\providecommand*\mcitethebibliography{\thebibliography}
\csname @ifundefined\endcsname{endmcitethebibliography}
  {\let\endmcitethebibliography\endthebibliography}{}

\end{document}